\documentclass[12pt,a4paper]{article}
\usepackage{amsfonts}
\usepackage{amsmath}
\usepackage{eurosym}
\usepackage{amssymb}
\usepackage{graphicx}
\usepackage{cite}
\usepackage[utf8]{inputenc}
\usepackage[colorlinks=true,urlcolor=blue,citecolor=blue]{hyperref}

\providecommand{\U}[1]{\protect\rule{.1in}{.1in}}
\input{tcilatex}
\begin{document}

\title{\textbf{Non-Hermitian Hamiltonian beyond }$\mathcal{PT}$\textbf{\
-symmetry for time-dependant }$SU(1,1)$\textbf{\ and }$SU(2)$\textbf{\
systems}\\
--- \textbf{exact solution and geometric phase in pseudo-invariant theory }}
\author{\textbf{Nadjat Amaouche}$^{a,1}$\textbf{, Maroua Sekhri}$^{a,2}$%
\textbf{, Rahma Zerimeche}$^{a\text{,}b,3}$\textbf{,} \and \textbf{Maamache
Mustapha}$^{a,4}$\textbf{\ and \ J.-Q. Liang}$^{c,5}${\small \thanks{$^{1}$%
nadjatamaouch@gmail.com,$^{2}$sekhrimaroua@gmail.com,$^{4}$%
zerimecherahma@gmail.com,$^{\text{4}}$maamache@univ-setif.dz, $^{5}$%
jqliang@sxu.edu.cn}} \\
%EndAName
$^{a}${\small Laboratoire de Physique Quantique et Syst\`{e}mes Dynamiques,}%
\\
{\small Facult\'{e} des Sciences, Ferhat Abbas S\'{e}tif 1, S\'{e}tif 19000,
Algeria.}\\
$^{b}${\small Physics Department, University of Jijel, BP 98, Ouled Aissa, }%
\\
{\small 18000 Jijel, Algeria.}\\
$^{c}${\small Institute of Theoretical Physics and Department of Physics,}\\
{\small State Key Laboratory of Quantum Optics and Quantum Optics Devices,}\\
{\small Shanxi University,Taiyuan, Shanxi 030006, China.}}
\date{}
\maketitle

\begin{abstract}
We investigate in this paper time-dependent non-Hermitian Hamiltonians,
which consist respectively of $SU(1,1)$ and $SU(2)$ generators. The former
Hamiltonian is PT symmetric but the latter one is not. A time-dependent
non-unitary operator is proposed to construct the non-Hermitian invariant,
which is verified as pseudo-Hermitian with real eigenvalues. The exact
solutions are obtained in terms of the eigenstates of the pseudo-Hermitian
invariant operator for both the $SU(1,1)$and $SU(2)$systems in a unified
manner. Then, we derive the LR phase, which can be separated to the dynamic
phase and the geometrical phase. The analytical results are exactly in
agreement with those of corresponding Hermitian Hamiltonians in the
literature.

\end{abstract}

\section{Introduction}

\hspace{0.38in}In conventional quantum mechanics the Hamiltonian is always
hermitian, this constrains the energy spectrum to be real . The
non-hermiticity means that the usual arguments for the reality of the
spectrum cannot be used. Carl Bender et al \cite{Bender98, Bender02}
interpreted the reality of this spectrum as being due to its $\mathcal{PT}$%
-symmety. That is, if we simultaneously reflect in space and reverse time,
the potential remains unchanged. Another idea that has been put forward as
an extension of conventional hermitian quantum mechanics, is
pseudo-hermiticity. The fact that it may be possible to find real
eigenvalues in a Hamiltonian which is non-hermitian was a point of interest,
particularly as the concept of $\mathcal{PT}$-symmety appears to have a more
physical interpretation than the very mathematical concept of hermiticity.
Then came the idea of the pseudo Hermiticity, which was a generalisation of
the conventional hermitian quantum mechanics. A Hamiltonian is said to be $%
\eta $\ pseudo-hermitian \cite{Schltz} if 
\begin{equation*}
\widehat{H}^{\dag }=\widehat{\eta }\widehat{H}\text{ }\widehat{\eta }^{-1}.
\end{equation*}%
Mostafazadeh \cite{Most1, Mosta2, Mostapha, mostapha} associated the
spectrum reality to a more general property than the $\mathcal{PT}$-symmety:
the pseudo-Hermiticity of the Hamiltonian. These new perspectives formed the
foundation of a new research field based on the fact that the Hermiticity is
a sufficient but unnecessary condition to have a real spectrum. Many
concrete Hamiltonian systems can not be described by autonomous Hamiltonians 
$\widehat{H}$, but require an explicit dependence on time $\widehat{H}(t)$.
In this work, we discuss how such type of systems can be treated
consistently when $\widehat{H}(t)$ is non-Hermitian  \cite{yuce, fr0,
fring1, Fring2, fr3}. While many researchers focused on studying
time-independent non-Hermitian systems. Others directed their efforts to
solving time dependent non-Hermitian systems in terms of the Lewis-
Riesenfeld (LR) invariant theory \cite{LewisR}, which presents the advantage
of obtaining exact solutions. Invariants are important in modern theoretical
physics and many theories are expressed in terms of their symmetries and
invariants. In particular, the invariants are capable of finding the
solution of the equation of motion. In the following, we recall the notion
of the pseudo-hermitien invariants introduced in \cite{KH, Maa17} which have
played a distinctive role in non hermitian quantum mechanics. In references 
\cite{KH, Maa17, kousa}, a particular attention was given to the special
subset of pseudo-Hermitian invariant operators associated to time dependent
non-Hermitian Hamiltonians, in which the reality of the eigenvalues of the
invariant is guaranteed. Let's review briefly the pseudo-Hermitian invariant
theory. The invariant operator $\widehat{I}(t)$ is said to be
pseudo-Hermitian with respect to the metric operator $\widehat{\eta }(t)$,
if 
\begin{equation*}
\widehat{I}^{\dag }(t)=\widehat{\eta }(t)\widehat{I}(t)\widehat{\eta }%
^{-1}(t),
\end{equation*}%
in which the metric operator is Hermitian. Thus the invariant $\widehat{I}(t)
$ can always be mapped to a Hermitian invariant operator $\widehat{I}^{h}(t)$
by a similarity (Dyson) transformation $\widehat{\rho }(t)$, such that%
\begin{equation*}
\widehat{I}^{h}(t)=\widehat{\rho }(t)\widehat{I}(t)\widehat{\rho }%
^{-1}(t)=\left( \widehat{I}^{h}(t)\right) ^{+},
\end{equation*}%
with $\widehat{\eta }(t)=\widehat{\rho }^{\dag }(t)\widehat{\rho }(t)$. The
exact solution of a $\mathcal{PT}$-symmetric non-Hermitian Hamiltonian was
presented recently for the periodically driven $SU(1,1)$ generators \cite%
{Yan, yen}. We emphasize that the spectrum reality of a non-Hermitian
Hamiltonian is not confined to the $\mathcal{PT}$-symmetry. In this paper,
we follow the pseudo-Hermitian invariant theory to solve the Schr\"{o}dinger
equation for a non-Hermitian Hamiltonian consisting of time-dependent $%
SU(1,1)$ and $SU(2)$ generators. The $SU(1,1)$ Hamiltonian is $\mathcal{PT}$
-symetric but the $SU(2)$ Hamiltonian is not. The paper is organized as
follows: in Sec.II we put forward a non-Hermitian Hamiltonian consisting of
periodically driven $SU(1,1)$ and $SU(2)$ generators. We propose a
non-unitary transformation operator $\hat{R}(t)$ to construct the
pseudo-Hermitian invariant.  A non-unitary transformation transforming the
time-dependent Schr\"{o}dinge equation of the free particle into that for
the quantum harmonic oscillator was considered in \cite{luis}. In Sec. III.
exact solutions of the Schr\"{o}dinger equation are found along with the LR
phase and non-adiabatic Berry phase, which reduces to the adiabatic phase in
slowly varying limit \cite{Ber}. The conclusion is given in Sec. IV.

\section{Non-Hermitian Hamiltonian and pseudo-Hermitian invariant}

The considered system is described by the following time dependent
Hamiltonian%
\begin{equation}
\hat{H}(t)=\Omega \hat{K}_{0}+G\left( \hat{K}_{+}\exp \left( i\phi
(t)\right) -\hat{K}_{-}\exp \left( -i\phi (t)\right) \right) ,  \label{Ham}
\end{equation}%
with $\Omega $, $G$ and $\phi (t)$ are real parameters: $\Omega $ being the
driving frequency, $G$ a coupling parameter and $\phi (t)=\omega t$ \ the
periodicity parameter $\phi (t)=\phi (t+T)$.

$\hat{K}_{0}$ is Hermitian, while $\left( \hat{K}_{-}\right) ^{+}=\hat{K}_{+}
$. These operators are $SU(1,1)$ and $SU(2)$ generators that satisfy these
commutation relations:%
\begin{equation}
\left\{ 
\begin{array}{c}
\left[ \hat{K}_{0},\hat{K}_{\pm }\right] =\pm \hat{K}_{\pm }, \\ 
\left[ \hat{K}_{+},\hat{K}_{-}\right] =D\hat{K}_{0}.%
\end{array}%
\right. ,  \label{commute}
\end{equation}%
where $D=\pm 2$ respectively for the $SU(2)$ and $SU(1,1)$ Lie algebras.

It is obvious that the Hamiltonian is periodic $\hat{H}(t)=\hat{H}(t+T)$ but
non Hermitian%
\begin{equation}
\hat{H}^{+}(t)=\Omega \hat{K}_{0}-G\left( \hat{K}_{+}\exp \left( i\phi
(t)\right) -\hat{K}_{-}\exp \left( -i\phi (t)\right) \right) \neq \hat{H}(t).
\label{H+}
\end{equation}

The non-Hermitian Hamiltonian Eq.(\ref{H+}) is $\mathcal{PT}$ -symetric for $%
SU(1,1$ system but is asymmetric for $SU(2)$. The time dependent Schr\"{o}%
dinger equation for this Hamiltonian is given by%
\begin{equation}
i\frac{\partial }{\partial t}\left\vert \psi (t)\right\rangle =\hat{H}%
(t)\left\vert \psi (t)\right\rangle .  \label{schro}
\end{equation}

Since the time-dependent Hamiltonian is not a conserved quantity, we solve
the time-dependent Schr\"{o}dinger equation (\ref{schro}) with the help of
the pseudo-Hermitian invariant scheme. The total time-derivative of the
invariant $\hat{I}(t)$ must be zero,%
\begin{equation}
i\frac{d\hat{I}(t)}{dt}=i\frac{\partial }{\partial t}\hat{I}(t)+\left[ \hat{I%
}(t),\hat{H}(t)\right] =0.  \label{cond}
\end{equation}%
\qquad We assume that the invariant $\hat{I}(t)$ can be generated from the
operator $\hat{K}_{0}$ such that%
\begin{equation}
\hat{I}(t)=\hat{R}(t)\hat{K}_{0}\hat{R}^{-1}(t),
\end{equation}%
where $\hat{R}(t)$ is a non-unitary transformation operator defined as%
\begin{equation}
\hat{R}(t)=\exp \left[ \frac{\varepsilon }{2}\left( \hat{K}_{+}\exp \left(
i\phi (t)\right) +\hat{K}_{-}\exp \left[ -i\phi (t)\right] \right) \right] ,
\end{equation}%
\begin{equation}
\hat{R}^{-1}(t)=\exp \left[ -\frac{\varepsilon }{2}\left( \hat{K}_{+}\exp
\left( i\phi (t)\right) +\hat{K}_{-}\exp \left[ -i\phi (t)\right] \right) %
\right] .
\end{equation}%
And $\varepsilon $ is a real parameter to be determined.

Using the following transformations \cite{Laiz, Laiy, Maama98} :%
\begin{eqnarray}
\hat{R}(t)\hat{K}_{+}\hat{R}^{-1}(t) &=&\hat{K}_{+}\cosh ^{2}\left( \frac{%
\alpha }{2}\right) -\hat{K}_{-}\exp \left( -2i\phi (t)\right) \sinh
^{2}\left( \frac{\alpha }{2}\right) -\hat{K}_{0}\exp \left( -i\phi
(t)\right) \sqrt{\frac{D}{2}}\sinh \left( \alpha \right) ,  \notag \\
\hat{R}(t)\hat{K}_{-}\hat{R}^{-1}(t) &=&\hat{K}_{-}\cosh ^{2}\left( \frac{%
\alpha }{2}\right) -\hat{K}_{+}\exp \left( 2i\phi (t)\right) \sinh
^{2}\left( \frac{\alpha }{2}\right) +\hat{K}_{0}\exp \left( i\phi (t)\right) 
\sqrt{\frac{D}{2}}\sinh \left( \alpha \right) ,  \notag \\
\hat{R}(t)\hat{K}_{0}\hat{R}^{-1}(t) &=&\hat{K}_{0}\cosh \left( \alpha
\right) -\frac{1}{\sqrt{2D}}\sinh \left( \alpha \right) \left( \hat{K}%
_{+}\exp \left( i\phi (t)\right) -\hat{K}_{-}\exp \left( -i\phi (t)\right)
\right) ,  \notag \\
i\hat{R}^{-1}(t)\frac{\partial }{\partial t}\hat{R}(t) &=&-2\omega \hat{K}%
_{0}\sinh ^{2}\left( \frac{\alpha }{2}\right) -\frac{\omega }{\sqrt{2D}}%
\sinh \left( \alpha \right) \left( \hat{K}_{+}\exp \left( i\phi (t)\right) -%
\hat{K}_{-}\exp \left( -i\phi (t)\right) \right) ,  \label{transf}
\end{eqnarray}

in which%
\begin{equation}
\alpha =\varepsilon \sqrt{\frac{D}{2}},
\end{equation}

we can obtain the invariant $\hat{I}(t)$ written as 
\begin{equation}
\hat{I}(t)=\hat{K}_{0}\cosh \left( \alpha \right) -\frac{1}{\sqrt{2D}}\sinh
\left( \alpha \right) \left( \hat{K}_{+}\exp \left( i\phi (t)\right) -\hat{K}%
_{-}\exp \left( -i\phi (t)\right) \right) .
\end{equation}

The invariant is obviously non-Hermitian, 
\begin{equation}
\hat{I}^{+}(t)=\hat{K}_{0}\cosh \left( \alpha \right) +\frac{1}{\sqrt{2D}}%
\sinh \left( \alpha \right) \left( \hat{K}_{+}\exp \left( i\phi \right) -%
\hat{K}_{-}\exp \left( -i\phi \right) \right) \neq \hat{I}(t).
\end{equation}

Substituting the invariant into the equation(\ref{cond}) we have 
\begin{equation*}
\left[ \hat{I}(t),\hat{H}(t)\right] =\left[ \frac{\Omega }{\sqrt{2D}}\sinh
\left( \alpha \right) +G\cosh \left( \alpha \right) \right] \left( \hat{K}%
_{+}\exp \left( i\phi \right) +\hat{K}_{-}\exp \left( -i\phi \right) \right),
\end{equation*}
and 
\begin{equation*}
i\frac{\partial }{\partial t}\hat{I}(t)=-\frac{\omega }{\sqrt{2D}}\sinh
\left( \alpha \right) \left( \hat{K}_{+}\exp \left( i\phi \right) +\hat{K}%
_{-}\exp \left( -i\phi \right) \right) .
\end{equation*}
The equation (\ref{cond}) is fulfilled under the auxiliary condition:%
\begin{equation}
G\cosh \left( \alpha \right) =-\frac{\omega +\Omega }{\sqrt{2D}}\sinh \left(
\alpha \right) ,  \label{auxi}
\end{equation}
from which the parameter $\varepsilon $ is determined. It is easy to check
that the invariant $\hat{I}(t)$ is pseudo-Hermitian with respect to the
metric operator $\widehat{\eta }$ 
\begin{equation}
\hat{I}^{\dag }(t)=\widehat{\eta }\hat{I}(t)\widehat{\eta }^{-1},
\end{equation}%
where%
\begin{equation}
\widehat{\eta }=\left( \hat{R}^{-1}\right) ^{\dag }\hat{R}^{-1}=\hat{R}^{-2},
\end{equation}%
for the Hermitian operator $\hat{R}^{\dag }=\hat{R}$.

\section{Exact solution and geometric phase}

As one of the most important results of the LR invariant theory, geometrical
phases attracted considerable interests in both theoretical and experimental
physics. The first general treatment of geometric phases is due to Berry 
\cite{Ber} who have considered Hermitian Hamiltonians undergoing adiabatic
changes. The instantaneous eigenstates returns to the same ray in the
Hilbert space, but acquires a phase factor consisting of a dynamical and a
geometrical part. Berry's phase knew many generalizations \cite{8}-\cite{11'}%
. There have also been attempts to extend geometric phases to systems
described by non-Hermitian Hamiltonians \cite{12}-\cite{24}.

Assuming that the pseudo-Hermitian invariant possesses a set of
non-degenerate eigenstates,%
\begin{equation}
\widehat{I}(t)\left\vert n(t)\right\rangle =\lambda _{n}\left\vert
n(t)\right\rangle ,
\end{equation}%
with the orthogonality condition 
\begin{equation*}
\left\langle n(t)\right\vert \widehat{\eta }(t)\left\vert m(t)\right\rangle
=\delta _{nm}.
\end{equation*}%
he general solution of the time-dependent Schr\"{o}dinger equation (\ref%
{schro}) is the superposition of the eigenstates of the pseudo-Hermitian
invariant $\hat{I}(t)$, 
\begin{equation}
\left\vert \psi \left( t\right) \right\rangle =\dsum_{n}C_{n}e^{i\alpha
_{n}(t)}\left\vert n(t)\right\rangle ,  \label{sol}
\end{equation}%
where $C_{n}$ are time independent coefficients and $\alpha _{n}(t)$ is the
LR phase. Substituting the general solution Eq.(\ref{sol}) into the Schr\"{o}%
dinger equation Eq.(\ref{schro}) yields the LR phase 
\begin{equation}
\alpha _{n}(t)=\int_{0}^{t}dt^{^{\prime }}\left\langle n\left( t^{^{\prime
}}\right) \right\vert \eta \left[ i\frac{\partial }{\partial t^{^{\prime }}}-%
\hat{H}(t^{^{\prime }})\right] \left\vert n\left( t^{^{\prime }}\right)
\right\rangle .
\end{equation}%
Using the transformation relations (\ref{transf}) and the auxiliary
condition (\ref{auxi}) we obtain the LR phase 
\begin{equation}
\alpha _{n}(t)=-\lambda _{n}\int_{0}^{t}dt^{^{\prime }}\left( \Omega +2\sqrt{%
\frac{D}{2}}G\sinh \left( \alpha \right) +2\left( \Omega +\omega \right)
\sinh ^{2}\left( \frac{\alpha }{2}\right) \right) ,  \label{tot}
\end{equation}%
in which 
\begin{equation}
\sinh ^{2}(\frac{\alpha }{2})=-\frac{1}{2}\pm \frac{(\omega +\Omega )}{2%
\sqrt{(\omega +\Omega )^{2}-2DG^{2}}}.\text{\ }  \label{sn}
\end{equation}%
The first term of LR phase $\alpha _{n}(t)$ gives rise to the geometrical
phase or Berry phase, which can be evaluated in one period of driving field $%
T=2\pi /\omega $ as 
\begin{subequations}
\begin{equation}
\gamma _{n}(T)=i\int_{0}^{T}dt^{^{\prime }}\left\langle n\left( t^{\prime
}\right) \right\vert \eta \frac{\partial }{\partial t^{\prime }}\left\vert
n\left( t^{\prime }\right) \right\rangle =-2\lambda _{n}\oint \sinh
^{2}\left( \frac{\alpha }{2}\right) d\phi .  \label{bery}
\end{equation}

Substituting the parameter of Eq.(\ref{sn}) into the Eq.(\ref{bery}) yields
the non-adiabatic Berry phase suitable for both $SU(2)$ and $SU(1,1)$
systems 
\end{subequations}
\begin{equation}
\mathcal{\gamma }_{n}(T)=-4\pi \lambda _{n}\left( -\frac{1}{2}\pm \frac{%
(\omega +\Omega )}{2\sqrt{(\omega +\Omega )^{2}-2DG^{2}}}\text{ \ }\right) .
\end{equation}

In the adiabatic approximation $\left( T\rightarrow \infty ,\dot{\phi}%
=\omega =0\right) $, the exact Berry-phase reduces to the well known
adiabatic form \cite{Yan} 
\begin{equation}
\mathcal{\gamma }_{n}(T)=-4\pi \lambda _{n}\left( -\frac{1}{2}\pm \frac{%
\Omega }{2\sqrt{\Omega ^{2}-2DG^{2}}}\right) .  \label{adiab}
\end{equation}

For $SU(1,1)$ system with $D=-2$, the generators of Lie algebra can be
expressed by boson creation and annihilation operators such that 
\begin{equation}
\hat{K}_{0}=\frac{1}{2}\left( \hat{a}^{\dag }\widehat{a}+\frac{1}{2}\right) ,%
\text{\qquad }\hat{K}_{+}=\frac{1}{2}\left( \hat{a}^{\dag }\right) ^{2},%
\text{\qquad }\hat{K}_{-}=\frac{1}{2}\left( \hat{a}\right) ^{2}.
\end{equation}

The non-Hermitian Hamiltonian is $\mathcal{PT}$-symmetric. The eigenstates
of $\hat{K}_{0}$ in this case are Fock states $\hat{a}^{\dag }\widehat{a}%
\left\vert n\right\rangle =n\left\vert n\right\rangle $ with the eigenvalues 
$\lambda _{n}=\frac{1}{2}\left( n+\frac{1}{2}\right) $,%
\begin{equation}
\hat{K}_{0}\left\vert n\right\rangle =\frac{1}{2}\left( n+\frac{1}{2}\right)
\left\vert n\right\rangle .  \label{vp}
\end{equation}

The LR phase (\ref{tot}) is then%
\begin{equation}
\alpha _{n}(t)=-\frac{1}{2}\left( n+\frac{1}{2}\right)
\int_{0}^{t}dt^{^{\prime }}\left( \Omega -G\sin \varepsilon +2\left( \Omega
+\omega \right) \sin \frac{\varepsilon }{2}\right) ,  \label{LR}
\end{equation}

in which 
\begin{equation*}
\sin ^{2}\frac{\varepsilon }{2}=-\frac{1}{2}\pm \frac{(\omega +\Omega )}{2%
\sqrt{(\omega +\Omega )^{2}+4G^{2}}}.
\end{equation*}

And the Berry phase in the adiabatic approximation (\ref{adiab}) is given by 
\begin{equation}
\mathcal{\gamma }_{n}(T)=-\pi \left( n+\frac{1}{2}\right) \left( -1\pm \frac{
\Omega }{\sqrt{\Omega ^{2}+4G^{2}}}\right) .
\end{equation}

For $D=2$ the generators of $SU(2)$ are simply spin operators. Since the
spin operators $\hat{K}_{0}$, $\hat{K}_{\pm }$ change respectively to $-\hat{
K}_{0}$, $-\hat{K}_{\mp }$ under $\mathcal{PT}$ transformation, the
non-Hermitian Hamiltonian is $\mathcal{PT}$ -symmetric. The eigenstates and
the eigenvalues of $\hat{K}_{0}$ are defined in this case as%
\begin{equation}
\hat{K}_{0}\left\vert j,n\right\rangle =n\left\vert j,n\right\rangle .
\end{equation}

The LR phase (\ref{tot}) in this case is,%
\begin{equation}
\alpha _{n}(t)=-n\int_{0}^{t}dt^{^{\prime }}\left( \Omega +G\sinh
\varepsilon +2\left( \Omega +\omega \right) \sinh ^{2}\frac{\varepsilon }{2}
\right) ,
\end{equation}

where 
\begin{equation*}
\sinh ^{2}(\frac{\varepsilon }{2})=-\frac{1}{2}\pm \frac{(\omega +\Omega )}{%
2 \sqrt{(\omega +\Omega )^{2}-4G^{2}}}
\end{equation*}

And the Berry phase in the adiabatic approximation (\ref{adiab}) is given by%
\begin{equation}
\mathcal{\gamma }_{n}(T)=-2\pi n\left( -1\pm \frac{\Omega }{\sqrt{\Omega
^{2}-4G^{2}}}\right) .
\end{equation}

\section{Conclusion}

\hspace{0.38in}It is well known that non-Hermitian Hamiltonians with $%
\mathcal{PT}$-symmetry can possess a real spectrum \cite{Bender98, Bender02}%
.However the spectrum reality is not restricted to the $\mathcal{PT}$%
-symmetry only. If a pseudo-Hermitian invariant exists for a non-Hermitian
Hamiltonian, the real spectrum is guaranteed. We solve the time-dependent
non-Hermitian Hamiltonian consisting of $SU(1,1)$ and $SU(2)$ generators
with the help of a pseudo-Hermitian invariant without considering the $%
\mathcal{PT}$- symmetry property. We propose a non-unitary but a Hermitian
transformation operator $\widehat{R}(t)$ to construct the non-Hermitian
invariant operator $\widehat{I}(t)$, which is proved to be pseudo-Hermitian
in regards to the metric operator given by $\widehat{\eta }=\widehat{R}^{-2}$%
. This invariant operator $\widehat{I}(t)$ possesses real eigenvalues for
both the $SU(1,1)$ and $SU(2)$ systems. Exact solutions are obtained in
terms of its eigenstates. We obtain the LR and the Berry phases, which are
in agreement with those of the corresponding Hermitian Hamiltonians in the
literature \cite{Laiy, Maama98}.

\end{document}